\documentclass[floats,preprint,prb,aps]{revtex4}
\usepackage{graphicx}
\begin{document}

\title{Superfluidity of ``dirty'' indirect magnetoexcitons  in coupled quantum wells
 in high magnetic field}

\author{Oleg L. Berman$^{1}$, Yurii E. Lozovik$^{2}$,  David W.
Snoke$^{3}$, and Rob D. Coalson$^{1}$}

\affiliation{\mbox{$^{1}$ Department of Chemistry, University of
Pittsburgh,}   \\ Pittsburgh, PA 15260, USA  \\
\mbox{$^{2}$ Institute of Spectroscopy, Russian Academy of
Sciences,}  \\ 142190 Troitsk, Moscow Region, Russia \\
\mbox{$^{3}$Department of Physics and Astronomy, University of
Pittsburgh,}  \\ Pittsburgh, PA 15260 USA  }

%\date{}

\begin{abstract}
Superfluidity in the quasi-two-dimensional  ($2D$)  system of
spatially indirect magnetoexcitons in coupled quantum wells (CQW)
and unbalanced two-layer electron system  in high magnetic field
$H$ is considered in the presence of a random field. The problem
of the rare gas of  magnetoexcitons with dipole-dipole repulsion
in a random field has been reduced to the problem of the rare gas
of dipole excitons without magnetic field
 with the effective magnetic mass of a magnetoexciton, which is
 a function of the magnetic field and parameters of the CQW, in an $H$-dependent effective random field.
 The density of the superfluid component $n_{s}$ and the temperature $T_{c}$
of the Kosterlitz-Thouless transition  to a superfluid state are
obtained as  functions of magnetic field $H$, interlayer
separation $D$ and the random field parameters $\alpha_{i}$ and
$g_{i}$. For $2D$ magnetoexcitonic systems, the rise of the
magnetic field $H$ and the interwell distance $D$ is found to
increase  the effective renormalized random field parameter $Q$
and suppress the superfluid density $n_s$ and the temperature of
the Kosterlitz-Thouless transition $T_c$. The suppressing
influence of $D$ on $n_s$ and $T_c$ in strong magnetic filed is
opposite to the case without magnetic field, when $n_s$ and $T_c$
increase with the rise of $D$ at fixed total exciton density $n$.
It is shown that in the presence of the disorder at sufficiently
large magnetic field $H$ or parameters of the disorder there is no
superfluidity at any exciton density.

\vspace{0.1cm}



\end{abstract}

\maketitle

\section{Introduction}

%\begin{large}

Systems of excitons with spatially separated electrons ($e$) and holes
 ($h$) (indirect excitons) in coupled  quantum wells (CQW) in
magnetic fields ($H$) have been the subject of
 several experimental investigations.
\cite{Chemla}$^-$\cite{Snoke} These systems without magnetic field
were also under the consideration
recently.\cite{Snoke_paper,Snoke_paper_Sc} They are of interest,
in particular, in connection with the possibility of superfluidity
of indirect excitons or $e-h$ pairs, which would manifest itself
in the CQW as  persistent electrical currents in each well
\cite{Lozovik}$^-$\cite{Berman_Willander}.  In high magnetic
fields ($H > 7 T$) two-dimensional (2D)
 excitons survive in a substantially wider temperature region, as the
exciton binding energies increase with magnetic field.
\cite{Lerner}$^-$\cite{Moskalenko} In addition, the 2D $e-h$
system in high $H$  fields is interesting due to the existence,
under some conditions of  supersymmetry in the system (for the
single quantum well)
 leading to
 unique exact solutions of the many-body problem (the last corresponding
to ideal Bose-condensation of magnetoexcitons at {\it any}
density)\cite{Lerner,Paquet}.

The superfluid state appears in the system under consideration below the
temperature of Kosterlitz-Thouless transition. \cite{Kosterlitz}  The
latter was studied recently for systems with spatially separated electrons
 ($e$) and holes ($h$) in the absence of magnetic field. \cite{Berman,Berman_Willander}

Attempts at experimental investigation of magnetoexciton
superfluidity in coupled quantum wells \cite{Chemla} make it
essential to study  the
 magnetic field dependence of the temperature of phase transition to the
superfluid state in systems of indirect magnetoexcitons   and to
analyze the density of the superfluid component.  It was shown
that increasing the magnetic field at a fixed magnetoexciton
density leads to  a lowering of the
 Kosterlitz-Thouless transition temperature $T_{c}$ on account of the
 increase of the exciton  magnetic mass as a function of a a magnetic field $H$ \ \cite{Berman_Tsvetus}.  But
it turns out that the  highest possible Kosterlitz-Thouless
transition temperature increases with  $H$ (at small $D$) due  to
the rise in the maximal density of magnetoexcitons {\it vs.} $H$ \
\cite{Berman_Tsvetus}.

The theory of the magnetoexcitonic systems did not take into
account the role of disorder, which is created by impurities and
boundary irregularities of the quantum wells. In real experiments,
however, disorder plays a very important role. Although the
inhomogeneous broadening linewidth of typical GaAs-based samples
has been improved from around 20 meV to less than 1
meV,\cite{Snoke_paper} the disorder energy is still not much
smaller than the exciton-exciton repulsion energy. At a typical
exciton density of $10^{10}$ cm$^{-2}$, the interaction energy of
the excitons is approximately $1$ meV.\cite{Snoke}  On the other
hand, the typical disorder energy of 1 meV is low compared to the
typical exciton binding energy of 5 meV.

A typical example of a two-dimensional system of weakly
interacting bosons is a system of indirect excitons in coupled
quantum wells
(GaAs/AlGaAs)\cite{Snoke_paper,Snoke_paper_Sc,Chemla,Krivolapchuk,Timofeev,Zrenner,Sivan}.
Fluctuations of the thickness of a quantum well, which arise
during the fabrication process,  impurities in the system, and
disorder in the alloy of the barriers can all lead to the
appearance of a random field. Of these, spectral analysis of the
exciton luminescence shows that alloy disorder, with a
characteristic length scale short compared to the excitonic Bohr
radius of around 100~\AA, plays the most important role.

The collective properties of 2D indirect excitons in weak disorder
without magnetic field were analyzed in
Ref.[\onlinecite{Berman_Ruvinsky}]. In
Ref.[\onlinecite{Berman_Snoke_Coalson}] we studied the case of a
random field which is not necessarily small compared to the
dipole-dipole repulsion between excitons for CQW without magnetic
field. The coherent potential approximation (CPA) allows us to
derive the 2D indirect exciton Green's function for a wide range
of the random field, resulting in the second order Born
approximation in weak scattering limit (the second order Born
approximation Green's function for 3D excitons was obtained by
Gevorkyan and Lozovik\cite{Gevorkyan}). It was predicted that in
the low-temperature limit, the density $n_{s}$ of the superfluid
component in CQW systems and the temperature of the superfluid
transition (the Kosterlitz-Thouless temperature
$T_{c}$\cite{Kosterlitz}) are the decreasing functions of the
random field\cite{Berman_Snoke_Coalson}.

In Ref.[\onlinecite{Ruvinsky_jetp}] the spectrum of the single
indirect 2D magnetoexciton (non-interacting with other
magnetoexcitons) in the strong perpendicular magnetic field in the
presence of disorder in the second-order Born approximation was
obtained.

In the present paper we derive the collective spectrum of the 2D
low-dense indirect magnetoexciton gas in the presence of the
disorder including the dipole-dipole repulsion between
magentoexcitons in the ladder approximation\cite{Berman_Tsvetus}.
We consider disorder which is not weak compared to the dipole
repulsion  in the second-order Born
approximation\cite{Ruvinsky_jetp}  The density of the superfluid
component $n_{s}$ and the temperature $T_{c}$ of the
Kosterlitz-Thouless transition to a superfluid state are obtained
as a function of magnetic field $H$, interlayer separation $D$ and
the random field parameters $\alpha_{i}$ and $g_{i}$. We show that
the density of the superfluid component and the
Kosterlitz-Thouless temperature of the superfluid phase transition
decrease as the random field increases. These results are derived
by the mapping the problem  of the rare indirect magnetoexciton
gas with the disorder in the strong magnetic field onto the
problem of the rare dipole gas without magnetic field consisting
of indirect excitons with the effective magnetic mass of a
magnetoexciton, which is a function of the magnetic field and the
parameters of the coupled quantum wells (CQW), in an $H$-dependent
effective random field. The application of these results, obtained
for CQW with spatially separated electrons and holes, for
unbalanced two-layer electron system is also discussed.

The paper is organized in the following way.  In Sec.~\ref{Sin_Ex}
we derive the effective Hamiltonian of rare ``dirty'' spatially
indirect magnetoexcitons in the effective magnetic mass
approximation in high magnetic field limit. In Sec.~\ref{single}
we obtain the Green's function of a single magnetoexciton in the
random field. In Sec.~\ref{col} the collective spectrum, the
superfluid density and the temperature of the Kosterlitz-Thouless
phase transition of magnetoexcitons are derived in the presence of
disorder. In Sec.~\ref{disc} we discuss our results and consider
the extension of the framework used for the indirect
magnetoexcitons in CQW on the system of indirect magnetoexcitons
in unbalanced two-layer electron system.

\section{The effective Hamiltonian of rare ``dirty'' spatially indirect magnetoexcitons in the effective magnetic mass
approximation}\label{Sin_Ex}

The total Hamiltonian $\hat H$ of 2D spatially separated $e$ and
$h$ in the perpendicular magnetic field in the presence of the
external field in the second quantization representation has the
form:
%-------------------------------------------------------------------------
\begin{eqnarray}\label{H_Tot}
\hat H &=& \int d \mathbf{R} \int d\mathbf{r}
\left[\hat{\psi}^{\dagger}(\mathbf{R},\mathbf{r})\left(\frac{1}{2m_{e}}
\left( -i \nabla_{e} + e\mathbf{A}_{e} \right)^2 +
\frac{1}{2m_{h}}\left( -i \nabla_{h} - e\mathbf{A}_{h} \right)^2
\right.\right. \nonumber \\ &-& \left.\left.\frac{e^2}{\epsilon
\sqrt{(\mathbf{r}_{e} - \mathbf{r}_{h})^{2} + D^{2}}} +
V_{e}(\mathbf{r}_{e}) +
V_{h}(\mathbf{r}_{h})\right)\hat{\psi}(\mathbf{R},\mathbf{r})\right]
\nonumber \\ &+& \frac{1}{2}\int d \mathbf{R}_{1} \int
d\mathbf{r}_{1}\int d \mathbf{R}_{2} \int d\mathbf{r}_{2}
\hat{\psi}^{\dagger}(\mathbf{R}_{1},\mathbf{r}_{1})\hat{\psi}^{\dagger}(\mathbf{R}_{2},\mathbf{r}_{2})
\left(U^{ee}(\mathbf{r}_{e1} - \mathbf{r}_{e2}) +
U^{hh}(\mathbf{r}_{h1} - \mathbf{r}_{h2})  \right. \nonumber
\\ &+& \left.
U^{eh}(\mathbf{r}_{e1} - \mathbf{r}_{h2}) + U^{he}(\mathbf{r}_{h1}
- \mathbf{r}_{e2})
\right)\hat{\psi}(\mathbf{R}_{2},\mathbf{r}_{2})\hat{\psi}(\mathbf{R}_{1},\mathbf{r}_{1}),
\end{eqnarray}
%-------------------------------------------------------------------------
Here $\hat{\psi}^{\dagger}(\mathbf{R},\mathbf{r})$ and
$\hat{\psi}(\mathbf{R},\mathbf{r})$ are the operators of creation
and annihilation of magnetoexcitons;
 $\mathbf{r}_{e}$ and $\mathbf{r}_{h}$ are electron and hole
locations along quantum wells, correspondingly; $\mathbf{A}_{e}$,
$\mathbf{A}_{h}$ are the vector potentials in electron and hole
location, correspondingly; $V_{e}(\mathbf{r}_{e})$ and
$V_{h}(\mathbf{r}_{h})$ represent the external fields acting on
electron and hole, correspondingly (we use units $c = \hbar = 1$);
 $D$ is the
distance between electron and hole quantum wells; $e$ is the
charge of an electron; $\epsilon$ is the dielectric constant. We
use below the coordinates of the magnetoexciton center of mass
$\mathbf{R} = (m_{e}\mathbf{r}_{e} + m_{h}\mathbf{r}_{h})/(m_{e} +
m_{h})$
 and the internal exciton coordinates
 $\mathbf{r} = \mathbf{r}_{e} - \mathbf{r}_{h}$.
The cylindrical gauge for vector-potential is used:
$\mathbf{A}_{e,h} = \frac{1}{2}\mathbf{H}\times \mathbf{r}_{e,h}$.
 $U^{ee}$, $U^{hh}$,
$U^{eh}$ and $U^{he}$ are the two-particle potentials of the
electron-electron, hole-hole, electron-hole and hole-electron
interaction, correspondingly, between electrons or holes from
different pairs:
%-------------------------------------------------------------------------
\begin{eqnarray}\label{Uee}
U^{ee}(\mathbf{r}_{e1} - \mathbf{r}_{e2}) &=&
\frac{e^{2}}{\epsilon |\mathbf{r}_{e1} - \mathbf{r}_{e2}|};
\hspace{0.5cm} U^{hh}(\mathbf{r}_{h1} - \mathbf{r}_{h2}) =
\frac{e^{2}}{\epsilon |\mathbf{r}_{h1} - \mathbf{r}_{h2}|} ;
\nonumber
\\ U^{eh}(\mathbf{r}_{e1} - \mathbf{r}_{h2}) &=& -
\frac{e^{2}}{\epsilon \sqrt{|\mathbf{r}_{e1} -
\mathbf{r}_{h2}|^{2} + D^{2}}}; \hspace{0.5cm}
U^{he}(\mathbf{r}_{h1} - \mathbf{r}_{e2}) = -
\frac{e^{2}}{\epsilon \sqrt{|\mathbf{r}_{h1} -
\mathbf{r}_{e2}|^{2} + D^{2}}}.
\end{eqnarray}
%-------------------------------------------------------------------------

The conserved quantity for isolated exciton in magnetic field
without any external field ($V_{e}(\mathbf{r}_{e}) =
V_{h}(\mathbf{r}_{h}) = 0$) (the exciton {\it magnetic} momentum)
is (see Ref.[\onlinecite{GorDzyal}]):
%-------------------------------------------------------------------------
\begin{equation}\label{Momentum}
\hat{\mathbf{P}} = -i \nabla _{e} -i \nabla _{h} + e
(\mathbf{A}_{e} - \mathbf{A}_{h}) - e\mathbf{ H} \times
(\mathbf{r}_{e} - \mathbf{r}_{h}) ,
\end{equation}
%-------------------------------------------------------------------------
The conservation of this quantity is  related to the invariance of
the system upon a simultaneous translation of $e$ and $h$ and
gauge transformation.

The eigenfunctions of Hamiltonian of a single isolated
magnetoexciton without any random field ($V_{e}(\mathbf{r}_{e}) =
V_{h}(\mathbf{r}_{h}) = 0$), which are also the eigenfunctions of
the magnetic momentum $\bf{P}$), have the following form (see
Refs.~[\onlinecite{Lerner,GorDzyal}]):
%-------------------------------------------------------------------------
\begin{eqnarray}\label{W_Func_Gen}
\Psi _{k\mathbf{P}} ({\bf R},{\bf r}) = \exp \left\{i {\bf R}
\left( {\bf P} + \frac{e}{2} {\bf H}\times {\bf R} \right) + i
\gamma \frac{\mathbf{P}\mathbf{r}}{2} \right\} \Phi_{k}
(\mathbf{P},\mathbf{r}) ,
\end{eqnarray}
%-------------------------------------------------------------------------
where $\Phi _{k} ({\bf P},{\bf r})$ is the function of internal
coordinates ${\bf r}$; ${\bf P}$ is the eigenvalue of magnetic
momentum; $k$ are quantum numbers of $j$ exciton internal motion.
In high magnetic fields
 $k = (n_{L},m)$, where  $n_{L} = min(n_{1}, n_{2})$, $m = |n_{1} - n_{2}|$, $n_{1(
2)}$ are Landau quantum numbers for $e$ and $h$
\cite{Lerner}$^,$\cite{Ruvinskiy}; $\gamma = (m_h - m_e)/(m_h +
m_e)$.

In this section we reduce the problem of
  the rare gas of  magnetoexcitons  (in high magnetic field) with dipole-dipole repulsion
in a random field  to the problem of the rare gas of dipole
excitons without magnetic field
 with the effective magnetic mass of a
magnetoexciton, which is a function of the magnetic field and
parameters of the quantum wells, in an $H$-dependent effective
random field. Our goal is to map the  Hamiltonian of the rare gas
of ``dirty'' dipole-dipole repulsing magnetoexcitons  placed on
the lowest Landau level (high magnetic field regime) onto the the
Hamiltonian of the rare gas of ``dirty'' dipole-dipole repulsing
excitons without the magnetic field, but characterized by the
effective magnetic mass dependent on $H$ and $D$ and the effective
random field, renormalized by magnetic field $H$.

We further expand the magnetoexciton field operators in a single
magnetoexciton basis set $\Psi _{k\mathbf{P}} ({\bf R},{\bf r})$
%---------------------------------------------------------------------
\begin{eqnarray}\label{expbs1}
 \hat{\psi}^{\dagger}(\mathbf{R},\mathbf{r}) = \sum_{k\mathbf{P}}
\Psi _{k\mathbf{P}}^{*} ({\bf R},{\bf
r})\hat{a}_{k\mathbf{P}}^{\dagger}; \hspace{1cm}
 \hat{\psi}(\mathbf{R},\mathbf{r}) = \sum_{k\mathbf{P}}
\Psi _{k\mathbf{P}}({\bf R},{\bf r})\hat{a}_{k\mathbf{P}},
\end{eqnarray}
%----------------------------------------------------------------------
where $\hat{a}_{k\mathbf{P}}^{\dagger}$ and
$\hat{a}_{k\mathbf{P}}$ are the corresponding  creation and
annihilation operators of a magnetoexciton in $(k,\mathbf{P})$
space.

We consider the case of strong magnetic field, when we neglect in
Eq.~(\ref{W_Func_Gen}) the transitions between different Landau
levels of the magnetoexciton caused by scattering by the slowly
changing in space potential $V_{e}(\mathbf{r}_{e}) +
V_{h}(\mathbf{r}_{h})$. We also neglect nondiagonal  matrix
elements of the Coulomb interaction between electron and hole in
the same pair. The application region of these two assumptions is
defined by the inequalities
%-------------------------------------------------------------------------
\begin{eqnarray}\label{ineq}
\omega_{c} \gg E_{b}, \hspace{2cm} \omega_{c} \gg
\sqrt{\left\langle V_{e(h)}^{2}\right\rangle _{av}},
\end{eqnarray}
%-------------------------------------------------------------------------
where $\omega_{c} = \sqrt{eH/m_{e-h}}$, $m_{e-h} =
m_{e}m_{h}/(m_{e} + m_{h})$ is the exciton reduced mass in the
quantum well plane; $E_{b}$ is the magnetoexciton binding energy
in an ideal ``pure'' system as a function of magnetic field $H$
and the distance between electron and hole quantum wells $D$:
$E_{b} \sim e^{2}/\epsilon r_{H}\sqrt{\pi/2}$ at $D \ll r_{H}$ and
$E_{b} \sim e^{2}/\epsilon D$ at $D \gg r_{H}$ ($r_{H} =
(eH)^{-1/2}$ is the magnetic length).\cite{Lerner,Ruvinskiy} Here
$\left\langle  \ldots \right\rangle _{av}$ denotes averaging over
the fluctuations of random field.

After exciton-exciton scattering their total magnetic momentum
$\mathbf{P} = \sum_{i}\mathbf{P}_{i}$ is conserved, but the
magnetic momentum $\mathbf{P}_{i}$ of each exciton can be changed.
Since in a strong magnetic field $\mathbf{\rho}$, the mean
distance between $e$ and $h$ along the quantum wells  is
proportional to the magnetic momentum (${\bf \rho} =
\frac{r_H^2}{H} \mathbf{H}\times \mathbf{P}$)
\cite{Lerner}$^,$\cite{GorDzyal}, the scattering is accompanied by
the exciton polarization.  We consider sufficiently low
temperatures when magnetoexciton states with only small magnetic
momenta $P \ll \frac{1}{r_{H}}$ are filled. The change of these
magnetic momenta due to exciton-exciton scattering  is also
negligible due to the conservation of the total magnetic momentum.
But these small magnetic momenta correspond to small separation
between electrons and holes along quantum wells $\rho \ll r_{H}$.
So magnetoexciton polarization due to scattering is negligible and
the magnetoexciton dipole moment keeps to be almost normal to
quantum wells $d = eD$, i.e. the interexciton interaction law is
not changed due to the scattering. For the rare system in the
strong magnetic field $n \ll r_{H}^{-2}$ ($n$ is the 2D density of
excitons; the characteristic radius of magnetoexciton along the
wells in the strong magnetic field at small $P$ approximately
equals to the magnetic length $r_{H}$\cite{Ruvinskiy}) the
exciton-exciton interaction is the dipole-dipole repulsion,
because the average distance between excitons $r_{s}$ is large
compare to the exciton radius ($r_{s} = (\pi n)^{-1/2} \gg
r_{H}$). So  for the rare system in the strong magnetic field at
$n \ll r_{H}^{-2}$ we have:
%-------------------------------------------------------------------------
\begin{eqnarray}\label{dipole_s}
\hat{U}(|\mathbf{R}_{1} - \mathbf{R}_{2}|) \equiv \hat{U}^{ee} +
\hat{U}^{hh} +  \hat{U}^{eh} + \hat{U}^{he} \simeq
\frac{e^{2}D^{2}}{\epsilon|\mathbf{R}_{1} - \mathbf{R}_{2}|^{3}}.
\end{eqnarray}
%-------------------------------------------------------------------------
Notice that the magnetoexcitons repel like parallel dipoles, and
their pair interaction potential depends only on the coordinates
of the center mass of the excitons and does not depend on the
coordinates of the relative motion of the electron and hole.

Now we substitute the expansions Eq.~(\ref{expbs1}) for the field
creation and annihilation operators in the total Hamiltonian
Eq.~(\ref{H_Tot}) and obtain the effective Hamiltonian in terms of
creation and annihilation operators in $\mathbf{P}$ space. In
strong magnetic fields
 $\omega _c = eH/\mu ^{*} \gg e^2/r_H$ the characteristic value  of $e-h$ separation in the
magnetoexciton $| \langle {\bf r} \rangle |$ has the order of the
magnetic length $r_H = 1/\sqrt{eH}$. The functions $\Phi _k ({\bf
P,r}) $ (see Eq.(\ref{W_Func_Gen}))
 are dependent on the difference $({\bf r - \rho})$,
where ${\bf \rho} = \frac{r_H^2}{H} \mathbf{H}\times \mathbf{P}$ \
\cite{Lerner}$^,$\cite{GorDzyal}.  At small magnetic momenta $P
\ll 1/r_H$ we have $\rho \ll r_H$, and, therefore, in functions
$\Phi _k ({\bf r-\rho}) $ we can ignore the variable ${\bf \rho}$
in comparison with ${\bf r}$. In  strong magnetic fields quantum
numbers $k$ correspond to the quantum numbers $(m,n_{L})$ (see
above). For the lowest Landau level we denote the spectrum of the
single exciton $\varepsilon _{0}(P) \equiv \varepsilon _{00}({\bf
P})$. In high magnetic field, when the typical interexciton
interaction $D^{2}n^{-\frac{3}{2}} \ll \omega _{c}$,
 one can ignore transitions between Landau levels and consider
only the states on the lowest Landau level $m=n_{L}=0$.  Using the
orthonormality of functions $\Phi _{mn} ({\bf 0},\mathbf{r})$ we
obtain the effective Hamiltonian $\hat{H}_{eff}$ in strong
magnetic fields. Since a typical value of $r$ is $r_H$, and $P \ll
1/r_H$ in this approximation the effective Hamiltonian
$\hat{H}_{eff}$ in the magnetic momentum representation $P$  on
the lowest Landau level $m=n_{L}=0$ has the same form  (compare
with Ref.[\onlinecite{Berman_Willander}])
 as for two-dimensional boson
system without a magnetic field, but with  the magnetoexciton
magnetic mass $m_{H}$ (which depends on $H$ and $D$) instead of
the exciton mass ($M = m_{e} + m_{h}$), magnetic momenta instead
of ordinary momenta and renormalized random field:
%-------------------------------------------------------------------------
\begin{eqnarray}\label{H_eff}
\hat{H}_{eff} &=& \sum_{\mathbf{P}}  \varepsilon_{0}(P)
\hat{a}_{\mathbf{P}}^{\dagger}\hat{a}_{\mathbf{P}}  +
\sum_{\mathbf{P},\mathbf{P}'}\left\langle
\mathbf{P}'\left|\hat{V}\right|\mathbf{P}\right\rangle
\hat{a}_{\mathbf{P}'}^{\dagger} \hat{a}_{\mathbf{P}} \nonumber
\\  &+&  \frac{1}{2} \sum_{\mathbf{P}_{1},\mathbf{P}_{2},\mathbf{P}_{3},\mathbf{P}_{4}}\left\langle
\mathbf{P}_{1},\mathbf{P}_{2}\left|\hat{U}\right|\mathbf{P}_{3},\mathbf{P}_{4}\right\rangle
\hat{a}_{\mathbf{P}_{1}}^{\dagger}
\hat{a}_{\mathbf{P}_{2}}^{\dagger}\hat{a}_{\mathbf{P}_{3}}\hat{a}_{\mathbf{P}_{4}},
\end{eqnarray}
%-------------------------------------------------------------------------
where $\hat{V} = \hat{V}_{e} + \hat{V}_{h}$. The dispersion
relation $\varepsilon _{0}(P)$ of an isolated magnetoexciton  on
the lowest Landau level is the quadratic function at small
magnetic momenta under consideration:
%-------------------------------------------------------------------------
\begin{equation}\label{Energy}
\varepsilon _{0}({\bf P}) \approx \frac{P^2}{2m_{H }},
\end{equation}
%-------------------------------------------------------------------------
where $m_{H }$ is the effective {\it magnetic} mass of a
magnetoexciton on the lowest Landau level, dependent on $H$ and
the distance $D$ between $e$ -- and $h$ -- layers (see
Ref.[\onlinecite{Ruvinskiy}]). In strong magnetic fields at $D \gg
r_{H}$ the exciton magnetic mass is $m_H \approx
D^{3}\epsilon/(e^{2}r_{H}^{4})$ \cite{Ruvinskiy}.  The quadratic
dispersion relation Eq.(\ref{Energy}) is true for small $P$ at
arbitrary magnetic fields $H$ and follows from the fact that $P =
0$ is an extremal point of the  dispersion law $\varepsilon
_{k}({\bf P})$. The last statement may be proved by taking into
account the regularity of the effective Hamiltonian $H_{{\bf P}}$
 as a function of the parameter ${\bf P}$ at ${\bf P} = 0$ and also the
invariance of $H_{{\bf P}}$ upon simultaneous rotation of ${\bf
r}$ and ${\bf P}$  in the CQW plane.
 For magnetoexciton ground state $m_{H} > 0$.
For high magnetic fields $r_{H} \ll a_{0}^{*}$  and at $D \lesssim
r_{H}$ the quadratic dispersion relation is valid at $P \ll
1/r_{H}$, but for $D \gg r_{H}$ it holds over a wider region
--- at least at $P \ll D/r_{H}^{2}$ \cite{Ruvinskiy} ($a_{0}^{*} =
\epsilon/(2m_{e-h} e^{2})$ is the radius of a 2D exciton at $H =
0$).

The matrix element of the inter-magnetoexciton interaction
$\left\langle
\mathbf{P}_{1},\mathbf{P}_{2}\left|\hat{U}\right|\mathbf{P}_{3},\mathbf{P}_{4}\right\rangle$
is defined as
%-------------------------------------------------------------------------
\begin{eqnarray}\label{dipole_mat_r}
\left\langle
\mathbf{P}_{1},\mathbf{P}_{2}\left|\hat{U}\right|\mathbf{P}_{3},\mathbf{P}_{4}\right\rangle
&=& \frac{1}{S^{2}}\int d^{2}R_{1}\int d^{2}R_{2}\int
d^{2}r_{1}\int d^{2}r_{2} U(\mathbf{R}_{1} - \mathbf{R}_{2})
\nonumber \\ && \Psi _{k_{1}\mathbf{P}_{1}}^{*} ({\bf R}_{1},{\bf
r}_{1})\Psi _{k_{2}\mathbf{P}_{2}}^{*} ({\bf R}_{2},{\bf
r}_{2})\Psi _{k_{3}\mathbf{P}_{3}} ({\bf R}_{1},{\bf r}_{1})\Psi
_{k_{4}\mathbf{P}_{4}} ({\bf R}_{2},{\bf r}_{2}) .
\end{eqnarray}
%-------------------------------------------------------------------------
 The matrix potential of $\hat{U}$ (Eq.~(\ref{dipole_s})) connecting the states
$\left\langle k_{1}=k_{2}=0,
\mathbf{P}_{1},\mathbf{P}_{2}\left|\right.\right.$ and
$\left\langle
k_{3}=k_{4}=0,\mathbf{P}_{3},\mathbf{P}_{4}\left|\right.\right.$
has the form
%-------------------------------------------------------------------------
\begin{eqnarray}\label{dipole_mat}
\left\langle
\mathbf{P}_{1},\mathbf{P}_{2}\left|\hat{U}\right|\mathbf{P}_{3},\mathbf{P}_{4}\right\rangle
= \frac{1}{S^{2}}U(\mathbf{P}_{3} -
\mathbf{P}_{1})\delta(\mathbf{P}_{1} + \mathbf{P}_{2} -
\mathbf{P}_{3} - \mathbf{P}_{4}),
\end{eqnarray}
%-------------------------------------------------------------------------
where $S$ is the area of a quantum well, and
%-------------------------------------------------------------------------
\begin{eqnarray}\label{dipole_fur}
U(\mathbf{P}_{3} - \mathbf{P}_{1}) = \int \int U(|\mathbf{R}_{1} -
\mathbf{R}_{2}|) \exp\left(i(\mathbf{P}_{3} -
\mathbf{P}_{1})(\mathbf{R}_{1} -
\mathbf{R}_{2})\right)d^{2}|\mathbf{R}_{1} - \mathbf{R}_{2}| .
\end{eqnarray}
%-------------------------------------------------------------------------

In the strong magnetic field limit, using for $\Phi _{k} ({\bf
P},\mathbf{r} - \mathbf{\rho})$ (${\bf \rho} = \frac{r_H^2}{H}
\mathbf{H}\times \mathbf{P}$) in Eq.~(\ref{W_Func_Gen}) the
internal wavefunction of the magnetoexciton at the lowest Landau
level,  at small magnetic momenta $P \ll 1/r_H$ we can ignore the
variable ${\bf \rho}$ relatively to ${\bf r}$ (see above). So in
$\Psi_{k=0,\mathbf{P}}(\mathbf{R},\mathbf{r})$
(Eq.~(\ref{W_Func_Gen})) we put $\mathbf{P} = 0$ and
$\mathbf{\rho} = 0$ for $\Phi _{k=0} ({\bf P},\mathbf{r} -
\mathbf{\rho})$, while we keep $\mathbf{P}\mathbf{R} \ne 0$ in the
exponent. This procedure is valid in a strong magnetic field at
small magnetic momenta, because the characteristic $\mathbf{\rho}$
is much smaller than the characteristic $R \sim r_{s} = (\pi
n)^{-1/2}$ ($r_{H} \ll (\pi n)^{-1/2}$). Keeping
$\mathbf{P}\mathbf{R} \ne 0$, we can use the magnetic momentum
conservation law below. So in the strong magnetic field limit,
using for $\Phi _{k} ({\bf 0},\mathbf{r})$ the internal
wavefunction of the magnetoexciton at the lowest Landau
level\cite{Ruvinskiy}
%-------------------------------------------------------------------------
\begin{eqnarray}\label{low_wf}
\Phi _{k=0} ({\bf 0},\mathbf{r}) =
\frac{1}{\sqrt{2\pi}r_{H}}\exp\left[ -
\frac{r^{2}}{4r_{H}^{2}}\right] ,
\end{eqnarray}
%-------------------------------------------------------------------------
we obtain the matrix element of the external potential
$V_{e,h}(\mathbf{r})$ connecting the states $\left\langle k=0,
\mathbf{P}\left|\right.\right.$ and $\left\langle
k=0,\mathbf{P}'\left|\right.\right.$, which is defined as
%-------------------------------------------------------------------------
\begin{eqnarray}\label{rand_mat_r}
\left\langle {\bf P}'\mid \hat{V}_{e,h}(\mathbf{r}) \mid {\bf
P}\right \rangle =  \frac{1}{S}\int d^{2}R\int d^{2}r
V_{e,h}(\mathbf{r}_{e,h}) \Psi _{k'\mathbf{P}'}^{*} ({\bf R},{\bf
r})\Psi _{k\mathbf{P}} ({\bf R},{\bf r}),
\end{eqnarray}
%-------------------------------------------------------------------------
and has the form
%-------------------------------------------------------------------------
$$
\left \langle{\bf P}'\mid \hat{V}_{e,h}(\mathbf{r}) \mid {\bf
P}\right \rangle=\frac{1}{S} \exp\left(-({\bf P}'-{\bf
P})^2\frac{r_{H}^2}{4}\right)
$$
$$
V_{e,h}({\bf P}'-{\bf P})\exp\left(\pm \frac{ir_{H}^2}{2H}{\bf
H}{\bf P}\times{\bf P}'\right)=
$$
\begin{equation}
\label{vpp} =\frac{1}{S} \tilde V_{e,h}({\bf P}'-{\bf P})
\exp\left(\pm \frac{ir_{H}^2}{2H}{\bf H}{\bf P}\times{\bf
P}'\right),
\end{equation}
%-------------------------------------------------------------------------
where
%-------------------------------------------------------------------------
\begin{eqnarray}\label{dipole_rand}
V_{e,h}({\bf P}'-{\bf P}) = \int \int V_{e,h} (\mathbf{r})
 \exp\left[i(\mathbf{P}' -
\mathbf{P})\mathbf{r}\right]d^{2}r.
\end{eqnarray}
%-------------------------------------------------------------------------

Then, using the expressions for the matrix elements
Eqs.~(\ref{dipole_mat_r}) and~(\ref{rand_mat_r}), the expansions
for the the field operators Eq.~(\ref{expbs1}), applying the
orthonormality of functions $\Phi _{k} ({\bf 0},\mathbf{r})$, and
employing only the lowest Landau level $k=0$ in the strong
magnetic field limit at the end we obtain from the
Eq.~(\ref{H_eff}) the effective Hamiltonian of dipole indirect
magnetoexctions in high magnetic field in the presence of the
disorder in coordinate space:
%-------------------------------------------------------------------------
\begin{eqnarray}\label{H_Tot_eff}
\hat H_{eff} &=& \int d \mathbf{R}
\hat{\psi}^{\dagger}(\mathbf{R})\left(-\frac{\nabla^2}{2m_{H}} +
V_{eff}(\mathbf{R})\right)\hat{\psi}(\mathbf{R}) \nonumber
\\ &+& \frac{1}{2}\int d \mathbf{R}_{1} \int d
\mathbf{R}_{2}
\hat{\psi}^{\dagger}(\mathbf{R}_{1})\hat{\psi}^{\dagger}(\mathbf{R}_{2})
U(\mathbf{R}_{1} - \mathbf{R}_{2})
\hat{\psi}(\mathbf{R}_{2})\hat{\psi}(\mathbf{R}_{1}),
\end{eqnarray}
%-------------------------------------------------------------------------
where $\hat{\psi}^{\dagger}(\mathbf{R})$ and
$\hat{\psi}(\mathbf{R})$ are the Bose creation and annihilation
field operators (the possibility of the assumption about
magnetoexcitons being bosons we discuss below), and the coupling
to the effective random field $V_{eff}(\mathbf{R})$ has the form
%-------------------------------------------------------------------------
\begin{eqnarray}\label{V_eff}
V_{eff}(\mathbf{R}) &=& \frac{1}{\pi r_{H}^{2}} \int \exp\left(
-\frac{(\mathbf{R}- \mathbf{r})^{2}}{r_{H}^{2}}\right) \left[
V_{e}(\mathbf{r}) + V_{h}(\mathbf{r}) \right] d \mathbf{r}.
\end{eqnarray}
%-------------------------------------------------------------------------
The Eq.(\ref{V_eff}) is valid, if the correlation length $L$ of
random potential ($V(\mathbf{r}_{e},\mathbf{r}_{h})$) is much
greater than the magnetoexciton mean size $r_{exc} = r_{H}$: ($L
\gg r_{H}$), i.e., in the case of a smooth surface potential,
which is realized, as was shown using a scanning tunnelling
microscope, on interfaces in $AlGaAs-GaAs$ structures discussed in
this paper. A more rigorous condition of the smoothness of random
potential can be expressed as\cite{Bonch}
%-------------------------------------------------------------------------
\begin{eqnarray}\label{rig}
r_{exc}\sqrt{\left\langle\nabla V^{2}\right\rangle_{av}} \ll
E_{b},
\end{eqnarray}
%-------------------------------------------------------------------------
and it holds for the strong magnetic field, when $r_{exc} = r_{H}
=(eH)^{-1/2}$, and $E_{b} \sim e^{2}/\epsilon D$ at $D \gg
r_{H}$\cite{Ruvinskiy,Ruvinsky_jetp}.

As a matter of fact, the effective magnetoexciton Hamiltonian
$\hat{H}_{eff}$ (Eq.~(\ref{H_Tot_eff})) treats the magnetoexciton
as an electrically neutral composite particle. Since the particle
is neutral, it does not directly interact with the magnetic field.
The interaction with the magnetic field manifests itself through
the renormalization of the exciton effective mass and modification
of the correlation function of the random field. Thus, we can do
mapping from the original problem of the rare weakly-interacting
magnetoexciton system in strong magnetic field in the presence of
the disorder, described by the total Hamiltonian
Eq.~(\ref{H_Tot}), on the rare system of excitons without magnetic
field  with the effective magnetic mass $m_{H}$ and in the
presence of the effective random field $V_{eff}$
(Eq.~(\ref{V_eff})), renormalized by the magnetic field $H$. The
dipole-dipole interaction term in the effective Hamiltonian
Eq.~(\ref{H_Tot_eff}) in the strong magnetic field limit is
exactly the same as one for the excitons without magnetic field,
because the dipole-dipole interaction does not depend on the
relative  coordinates of electron and hole in the exciton as long
as we assume the excitons are parallel dipoles (as mentioned
above). This mapping allows us to use the results  for the
collective spectrum, superfluid density and Kosterlitz-Thouless
temperature, obtained for the system in random field without
magnetic field \cite{Berman_Snoke_Coalson}, for the case of strong
magnetic field.

\section{The Green's function of a single magnetoexciton in the random field}
\label{single}

The interaction between an spatially indirect exciton in coupled
quantum wells and a random field, induced by fluctuations in
widths of electron and hole quantum wells, has the form
\cite{Ruvinsky_jetp}:
%-------------------------------------------------------------------------
\begin{eqnarray}
\label{gaussian} V(\mathbf{r}_{e},\mathbf{r}_{h}) =
\alpha_{e}[\xi_{1}(\mathbf{r}_{e}) - \xi_{2}(\mathbf{r}_{e})] +
\alpha_{h}[\xi_{3}(\mathbf{r}_{h}) - \xi_{4}(\mathbf{r}_{h})] ,
\end{eqnarray}
%------------------------------------------------------------------------
where $\alpha_{e,h} = \partial E_{e,h}^{(0)}/\partial d_{e,h}$,
$d_{e,h}$ is the average  widths of the electron and hole quantum
wells, $E_{e,h}^{(0)}$ are the lowest levels of the electron and
hole in the conduction and valence bands, and $\xi_{1}$ and
$\xi_{2}$ ($\xi_{3}$ and $\xi_{4}$) are fluctuations in the widths
of the electron (hole) wells on the upper and lower interfaces,
respectively. We assume that fluctuations on different interfaces
are statistically independent, whereas fluctuations of a specific
interface are characterized by Gaussian correlation function
%-------------------------------------------------------------------------
\begin{eqnarray}
\label{gaussian1} \left\langle
\xi_{i}(\mathbf{r}_{1})\xi_{j}(\mathbf{r}_{2}) \right\rangle =
g_{i}\delta_{ij}\delta(\mathbf{r}_{2} - \mathbf{r}_{1}),
\hspace{0.5in} \left\langle \xi_{i}(\mathbf{r}) \right\rangle = 0,
\end{eqnarray}
%-------------------------------------------------------------------------
where $g_{i}$ is proportional to the squared amplitude of the
$i$th interface fluctuation\cite{Ruvinsky_jetp}. This is possible
if the distance $D$ between the electron and hole quantum wells is
larger than the amplitude of fluctuations on the nearest surfaces.

We consider the characteristic length of the random field
potential  $L$ to be much shorter than the average distance
between excitons $r_{s} \sim 1/\sqrt{\pi n}$  ($L \ll 1/\sqrt{\pi
n}$, where $n$ is the total exciton density). Therefore, in order
to obtain the Green's function of the magnetoexcitons with
dipole-dipole repulsion in the random field,  we obtain the
Green's function of a single magnetoexciton in the random field
(not interacting with other magnetoexcitons), and then apply the
perturbation theory with respect to dipole-dipole repulsion
between exciton, using the system of the non-interacting
magnetoexcitons as a reference system in analogy with the system
without magnetic field\cite{Berman_Snoke_Coalson}.

Since the effective magnetoexciton Hamiltonian
Eq.~(\ref{H_Tot_eff}) is translationally invariant, we can write
the Green function of isolated magnetoexciton in the momentum
space. The Green's function of the center of mass of the isolated
magnetoexciton at $T = 0$ in the momentum-frequency domain
($G^{(0)}(\mathbf{p}, \omega)$) in the random field in the
coherent potential approximation (CPA) is given
by\cite{Ruvinsky_jetp} (here and below $\hbar = 1$)
%-------------------------------------------------------------------------
\begin{eqnarray}
\label{green_0} G^{(0)}(\mathbf{p}, \omega) = \frac{1}{\omega -
\varepsilon_{0}(p) + \mu + i Q(\mathbf{p}, \omega) }  ,
\end{eqnarray}
%-------------------------------------------------------------------------
where $\mu$ is the chemical potential of the system, and
$\varepsilon_{0}(p) = p^{2}/2m_{H}$ is the spectrum of the center
mass of the exciton in the ``clean'' system. The function
$Q(\mathbf{p}, \omega)$ is determined by effective random field
acting on the center of mass of the exciton. For zero random
field, $Q(\mathbf{p}, \omega) \rightarrow 0$. If $g_{i} m_{H} \ll
E_{b}$ the function $Q(\mathbf{p}, \omega)$ in the coherent
potential approximation is given by\cite{Ruvinsky_jetp}
%-------------------------------------------------------------------------
\begin{eqnarray}
\label{Q_def1} Q (\mathbf{p}, \omega) = \frac{1}{2} \int
G^{(0)}(\mathbf{q}, \omega) B(|\mathbf{p} - \mathbf{q}|)
\frac{d^{2}q}{(2\pi)^{2}}  ,
\end{eqnarray}
%-------------------------------------------------------------------------
where
%----------------------------------------------------------------------
\begin{equation}\label{fourier}
   B(\mathbf{p}) \equiv  \int d^{2} R B(\mathbf R) e^{- i \mathbf{p R}},
\end{equation}
%----------------------------------------------------------------------
for $\mathbf{R} = \mathbf{R}_{1} -  \mathbf{R}_{2}$, and in the
coordinate domain $B(\mathbf{R}_{1},\mathbf{R}_{2}) = \left\langle
V_{eff}(\mathbf{R}_{1})V_{eff}(\mathbf{R}_{2})\right\rangle_{av}$
(the effective potential $V_{eff}$ is given by Eq.~(\ref{V_eff}))
has the form \cite{Ruvinsky_jetp}
%----------------------------------------------------------------------
\begin{eqnarray}\label{B_def}
      B(\mathbf{R}_{1},\mathbf{R}_{2}) = B(\mathbf{R}_{1} - \mathbf{R}_{2}) =
      \frac{\alpha_{e}^{2}(g_{1} + g_{2}) + \alpha_{h}^{2}(g_{3} + g_{4})}{2\pi r_{H}^{2}}
     \exp\left( - \frac{(\mathbf{R}_{1} - \mathbf{R}_{2})^{2}}{2r_{H}^{2}}\right).
\end{eqnarray}
%----------------------------------------------------------------------
Note, that, since in the limit of  strong magnetic fields the
magnetic length $r_{H} = (eH)^{-1/2}$ is much smaller than the
characteristic  length of the random field potential $L$ ($r_{H}
\ll L$), and, therefore,
%----------------------------------------------------------------------
\begin{eqnarray}\label{B_def1}
     \lim_{r_{H}|\mathbf{R}_{1} - \mathbf{R}_{2}|^{-1}\rightarrow 0} B(\mathbf{R}_{1} - \mathbf{R}_{2}) =
      \frac{\alpha_{e}^{2}(g_{1} + g_{2}) + \alpha_{h}^{2}(g_{3} + g_{4})}{\sqrt{2\pi}
      r_{H}}\delta(\mathbf{R}_{1} - \mathbf{R}_{2}),
\end{eqnarray}
%----------------------------------------------------------------------
the random field acting on a magnetoexciton is represented by the
white noise correlation function of the random potential
$B(\mathbf{R}_{1},\mathbf{R}_{2}) = \left\langle
V_{eff}(\mathbf{R}_{1})V_{eff}(\mathbf{R}_{2})\right\rangle_{av}$.
 The time-reversal symmetry of the effecive magnetoexciton Schr\"{o}dinger equation in strong magnetic field,
 corresponding to
  the effective Hamiltonian Eq.~(\ref{H_eff}) with the effective
  random field Eq.~(\ref{V_eff}),
and elimination of the long-range property of the random field
(Eq.~(\ref{B_def1})) cancels all effects related with the broken
time-reversal symmetry in the Schr\"{o}dinger equation
Eq.~(\ref{H_Tot}) at $m_{e} \neq m_{h}$, that result in
corrections to the Green's function of a
magnetoexciton.\cite{Arseyev}

Using Eq.~(\ref{fourier}), we obtain the Fourier transform of
$B(R)$
%-------------------------------------------------------------------------
\begin{eqnarray}
\label{B_p}  B(p) = \frac{\alpha_{e}^{2}(g_{1} + g_{2}) +
\alpha_{h}^{2}(g_{3} + g_{4})}{16\pi^{4}}\exp\left( -
\frac{r_{H}^{2}p^{2}}{32}\right) .
\end{eqnarray}
%-------------------------------------------------------------------------

So the CPA Green's function of the 2D indirect exciton is
determined by the solution of the self-consistent equations
Eqs.~(\ref{green_0}) and~(\ref{Q_def1}).

In the weak-scattering limit (~$g_{i} m_{H}\ll E_{b} \sim
e^{2}/\epsilon D$~) we use the second-order Born approximation for
$Q$ similar to
Refs.~[\onlinecite{Gevorkyan,Ruvinsky_jetp,Berman_Snoke_Coalson}],
expanding $Q$ (Eq.~(\ref{Q_def1})) in series to the first order in
$B(|\mathbf{p} - \mathbf{q}|)$ (which is the first order in
$g_{i}$), and we replace Eq.~(\ref{Q_def1}) by:
%-------------------------------------------------------------------------
\begin{eqnarray}
\label{Q_def} Q (\mathbf{p}, \omega) = \frac{\pi}{2} \int \delta
\left(\omega -  \frac{q^{2}}{2m_{H}}\right) B(|\mathbf{p} -
\mathbf{q}|) \frac{d^{2}q}{(2\pi)^{2}} .
\end{eqnarray}
%-------------------------------------------------------------------------
Substituting $B(p)$ from Eq.~(\ref{B_p}) into Eq.~(\ref{Q_def}),
we obtain for $Q (\mathbf{p}, \omega)$
%-------------------------------------------------------------------------
\begin{eqnarray}
\label{Q_res}  Q (\mathbf{p}, \omega) &=&
\frac{\alpha_{e}^{2}(g_{1} + g_{2}) + \alpha_{h}^{2}(g_{3} +
g_{4})}{64\pi^{4}}m_{H}\exp\left( -\frac{r_{H}^{2}}{32} (p^{2} +
2m_{H}\omega)\right) \nonumber \\ && J_{0}\left(
\frac{r_{H}^{2}}{16} \sqrt{2m_{H}\omega}p \right),
\end{eqnarray}
%-------------------------------------------------------------------------
where $J_{0}(z)$ is the Bessel first integral. The second-order
Born Green function of the single indirect exciton in the
  random field in CQW $G^{(0)}(\mathbf{p}, \omega)$ is derived by
substituting $Q (\mathbf{p}, \omega)$ from
  Eq.~(\ref{Q_res}) into Eq.~(\ref{green_0}).

\section{Collective spectrum and superfluidity of indirect dirty magnetoexcitons}
\label{col}

Due to the interwell separation  $D$, indirect magnetoexcitons
both in the ground state and in excited states have electrical
dipole moments. We suppose  that indirect excitons interact as
{\it parallel} dipoles.
 This is valid
when $D$ is larger than the mean separation $\langle  r \rangle $
between electron and hole in the magnetoexciton along quantum
wells $D \gg \langle r \rangle$. We take into account that at high
magnetic fields $\langle  r \rangle \approx Pr_H^2 $ ($\langle
{\bf r} \rangle $ is normal to ${\bf P}$)
 and that the typical value of magnetic
momenta (with exactness to logarithm of the exciton density
($ln(n_{ex})$, see below) is $P \sim \sqrt{n_{ex}}$ (if the dispersion
relation $\varepsilon _{k} (P) = \frac{P^2}{2m_{H k}}$ is true).  So the
inequality $D \gg \langle r \rangle$ is valid at $D \gg \sqrt{n}
r_H^2 $.

The distinction between excitons and bosons manifests itself in
exchange effects (see
Refs.~[\onlinecite{Keldysh},\onlinecite{Berman}]). These effects
for excitons with spatially separated $e$ and $h$ in a rare system
$n_{ex}a^{2}(H,D) \ll 1$ are suppressed due to the negligible
overlapping of wave functions of two excitons on account of the
potential barrier, associated with the dipole-dipole repulsion of
indirect excitons \cite{Berman} (here $n_{ex}$, $a(D,H) $ are
respectively
 density and magnetoexciton radius along quantum
wells, respectively). The small tunnelling parameter connected
with this barrier is $exp[- \int_{a(H,D)}^{r_{0}}\sqrt{2m_{H
k}(\frac{e^{2}D^{2}}{R^{3}} - \frac{\kappa ^{2}}{2m_{H k}})} dR]$,
where $\kappa ^{2} \sim \frac{n}{\ln \left( \frac{1}{8\pi n m_{H
k}^2 e^4 D^4} \right)}$ is the characteristic momentum of the
system (see below); $r_{0} = (2m_{H k}e^{2}D^{2}/\kappa
^{2})^{1/3}$ is the classical turning point for the dipole-dipole
potential at the energy equal to the chemical potential of the
system. In high magnetic fields the small parameter mentioned
above has the form $\exp[-2(m_{H k})^{1/2}eD a^{-1/2}(H,D)]$.
Therefore the system of indirect magnetoexcitons can be treated by
the diagram technique for a boson system.

 The rare excitonic  system    is stable
 at $D < D_{cr}(H)$ and $T = 0$ when the magnetoexciton energy $E_{b}(D,H)$
(calculated in Ref.[\onlinecite{Ruvinskiy}]) is larger than the
sum of activation energies $E_{L} = k\frac{e^{2}}{\epsilon r_{H}}$
for
 incompressible
Laughlin
  liquids of electrons or holes;  $k = 0.06$
for $\nu = \frac{1}{3}$ etc. \cite{Hall}. Since $k \ll 1$, the
critical value $D_{cr} \gg r_{H}$. In this case one has $E_{b} =
\frac{e^{2}}{\epsilon D}(1 - \frac{r_{H}^{2}}{D^{2}})$ for a
magnetoexciton with quantum numbers $n = m = 0$ (see
Ref.[\onlinecite{Ruvinskiy}]). As a result we have from the
stability condition (see above)
 $D_{cr} = r_{H} (\frac{1}{2k} - 2k)$.
 For greater $\nu $  it gives an upper bound on $D_{cr}$.
  In the rare system in high magnetic
 fields $\mu \ll E_{b}$ (see below).
The coefficient $k$ in the activation energy $E_{L}$ may be
represented as $k = k_{0}\sqrt{\nu }$. So from the relation
between $D_{cr}$ and $r_{H}$ one has: $\nu _{cr} =
\frac{1}{k_{0}^{2}} \frac{r_{H}^{2}}{D^{2}} (1 -
\frac{r_{H}^{2}}{8D^{2}})$. Thus maximal density for stable
magnetoexciton phase is $n_{max} = \nu _{cr}/2\pi r_{H}^{2}$ .

 At $\nu =
\frac{1}{2}$ the elecron-hole phase  can be unstable  due to the
pairing of electron and hole composite fermions (which form the
Fermi surface of composite fermions in the mean field
approximation \cite{HLR}).

Since the effective Hamiltonian  $\hat{H}_{eff}$
(Eq.~(\ref{H_Tot_eff})) of the system of indirect ``dirty''
magnetoexcitons at small momenta  is exactly identical to the
Hamiltonian of indirect ``dirty'' excitons without magnetic field
replacing the excitonic mass $M = m_{e} + m_{h}$ by the magnetic
mass $m_{H}$ and using the effective random field $V_{eff}$, we
can use the expressions for the ladder approximation
 Green's function\cite{Abrikosov,Yudson}, collective spectrum, normal and superfluid
 density and the temperature of Kosterlitz-Thouless phase
 transition \cite{Kosterlitz} for the ``dirty'' system without magnetic field\cite{Berman_Snoke_Coalson} replacing excitonic mass and
 $Q$. Since
the characteristic length of the random field potential  $L$ is
much shorter than the average distance between electron and hole,
the first step in the averaging procedure with respect to the
random field results in the Green function of a non-interacting
exciton with the imaginary part. And the second step takes care of
the exciton-exciton repulsion. The results of
Ref.~[\onlinecite{Berman_Snoke_Coalson}] show the density of the
superfluid component and the temperature of the phase transition
decrease with the rise of the disorder, which give the good
correspondence with the results of the Lopatin-Vinokur approach
for the weakly interacting bosons in the presence of
disorder.\cite{Lopatin}

Since the characteristic frequencies and momenta which give the
greatest contribution to the Green's function in the ladder
approximation are\cite{Yudson} $\omega \epsilon D/e^{2} \sim
n/\log[r_{H}^{2}/(nD^{4})] \ll 1$ and $p r_{H} \sim m_{H}
\sqrt{n/\log[r_{H}^{2}/(nD^{4}))]} \ll 1$, respectively, for the
single exciton Green's function $G^{(0)} (\mathbf{p}, \omega)$,
participating in the ladder approximation, we approximate $Q
(\mathbf{p}, \omega)$ by $Q (\mathbf{p} = \mathbf{0}, \omega = 0)$
(see Eq.~(\ref{Q_res}))
%-------------------------------------------------------------------------
\begin{eqnarray}
\label{Q_res_ap}  Q (\mathbf{p}, \omega) &=& Q =
\frac{\alpha_{e}^{2}(g_{1} + g_{2}) + \alpha_{h}^{2}(g_{3} +
g_{4})}{64\pi^{4}}m_{H}.
\end{eqnarray}
%-------------------------------------------------------------------------
The chemical potential $\mu $ is obtained in the form (compare to
Ref.[\onlinecite{Berman_Snoke_Coalson}]):
%-------------------------------------------------------------------------
\begin{equation}\label{Mu}
\mu =  \frac{\kappa ^2 }{2m_{H}} = \frac{8\pi n}{2m_{H} \log
\left( \frac{\epsilon^{2}}{8\pi n m_{H}^2 e^4 D^4} \right)} .
\end{equation}
%-------------------------------------------------------------------------
where $\kappa$ is a characteristic momentum. We have the
condensate Green's function $D(\mathbf{p}, i\omega_{k})$ (compare
to Ref.[\onlinecite{Berman_Snoke_Coalson}]):
%-------------------------------------------------------------------------
\begin{eqnarray}
\label{d0} D^{(0)}({\bf p},i\omega _{k}) = \frac{- i(2\pi )^{2}
n_{0}\delta ({\bf p})}{i\omega _{k} + \mu + i Q},
\end{eqnarray}
%-------------------------------------------------------------------------
where $n_{0}$ is the density of Bose condensate. Since at small
temperatures \\ $(n - n_{0})/n \ll 1$, according to the ladder
approximation\cite{Abrikosov} we use below $n$ instead of $n_{0}$.
$G(\mathbf{p}, i\omega_{k})$ and $F(\mathbf{p}, i\omega_{k})$ are
the normal and anomalous Green functions of the overcondensate:
%-------------------------------------------------------------------------
\begin{eqnarray}
\label{g0} G({\bf p},i\omega _{k}) &=& - \frac{i \omega _{k} +
\varepsilon _{0}(p)
  + \mu + i Q}{
\omega _{k}^{2} + \varepsilon ^{2}(p) - 2i (\mu - \varepsilon
_{0}(p))Q}  ; \nonumber\\
  F({\bf
p},i\omega _{k}) &=&  - \frac{\mu }{ \omega _{k}^{2} + \varepsilon
^{2}(p) - 2i(\mu - \varepsilon _{0}(p)) Q} ,
\end{eqnarray}
%-------------------------------------------------------------------------
where $\varepsilon _{0}(p)$ is the spectrum of noninteracting
excitons; the spectrum of interacting excitons  has the form
$\varepsilon (p) = \sqrt{\left(p^{2}/(2m_{H}) + \sqrt{\mu^{2} -
Q^{2}}\right)^{2} - (\mu^{2} - Q^{2})}$, and for small momenta $p
\ll \sqrt{2m_{H}\mu}$ the excitation spectrum is acoustic
$\varepsilon (p) = c_{s} p$, where $c_{s} = \sqrt{\sqrt{\mu^{2} -
Q^{2}}/m_{H}}$ is the velocity of sound.

At large magnetic momenta $P$ the isolated magnetoexciton spectrum
$\varepsilon (P)$ contrary to the case $H = 0$ has a constant
limit (being equal to Landau level $\frac{\omega _{c}}{2}$ for
reduced effective mass, see Refs.[\onlinecite{Ruvinskiy}],
[\onlinecite{Lerner}]).  As
 a result the spectrum of the  interacting magnetoexciton
system also has a plateau at high momenta. Therefore, the Landau
criterium of superfluidity is not strictly valid at large $P$ for
the interacting magnetoexciton system. However, the probability of
excitation of quasiparticles  with magnetic momenta $P \gg
1/r_{H}$ by moving magnetoexciton system is negligibly small at
small superfluid velocities\cite{Berman_Tsvetus}. In this sense,
the appearance of a linear branch can be taken as the criterion
for superfluidity of 2D magnetoexcitons.

 The density of the superfluid
component $n_{s}(T)$ can be obtained using the relation $n_{s}(T)
= n - n_{n}(T)$, where $n_{n}(T)$ is the density of the normal
component. The density of normal component $n_{n}$ is (compare
to\cite{Berman_Snoke_Coalson}):
%-------------------------------------------------------------------------
\begin{eqnarray}
\label{nn4} n_{n} = n_{n}^{0} + \frac{N}{m_{H}}
\int_{}^{}\frac{d{\bf p}}{(2\pi )^{2}} p^{2}\mu \frac{\varepsilon
_{0}(p)}{ \varepsilon ^{4}(p)}Q       .
\end{eqnarray}
%-------------------------------------------------------------------------
Here $N$ is the total number of particles, and $n_{n}^{0}$ is the
density of the normal component in a pure system with no
impurities:
%-------------------------------------------------------------------------
\begin{eqnarray}
\label{Bose} n_{n}^{0}  = - \frac{1}{2m_{H}} \int_{}^{}
\frac{d{\bf p}}{(2\pi )^{2}} p^{2} \frac{\partial
n_{0}(p)}{\partial \varepsilon}.
\end{eqnarray}
%-------------------------------------------------------------------------
where $n_{0}(p)=(e^{\varepsilon (p)/T} - 1)^{-1}$  is the
distribution of an ideal Bose gas of temperature excitations.

The first term in Eq.~(\ref{nn4}), which does not depend on $Q$,
is the contribution to the normal component due to scattering of
quasiparticles with an acoustic spectrum in a pure system at $T
\neq 0$ (compare with Ref.~[\onlinecite{Berman_Snoke_Coalson}]),
%-------------------------------------------------------------------------
\begin{eqnarray}
\label{nn00} n_{n}^{0} =  \frac{3 \zeta (3) }{2 \pi }
\frac{T^3}{c_{s}^{4}(n,Q) m_{H}},
\end{eqnarray}
%-------------------------------------------------------------------------
where $\zeta (z)$ is the Riemann zeta function ($\zeta (3) \simeq
1.202$).
  The second term in Eq.~(\ref{nn4})
is the contribution to the normal component due to the interaction
of the particles (magnetoexcitons) with the random field,
%-------------------------------------------------------------------------
\begin{eqnarray}
\label{nn55} n_{n} = n_{n}^{0} + \frac{n Q}{2m_{H}c_{s}^{2}(n,Q)}
= \frac{3 \zeta (3) }{2 \pi } \frac{T^3}{c_{s}^{4}(n,Q) m_{H}} +
\frac{n Q}{2m_{H}c_{s}^{2}(n,Q)}.
\end{eqnarray}
%-------------------------------------------------------------------------
The density of the superfluid component is $n_{s} = n - n_{n}$.
 From Eqs.~(\ref{nn00}) and~(\ref{nn55}) we can see, that the
random field decreases the density of the superfluid component.

In a 2D system, superfluidity appears below the
Kosterlitz-Thouless transition temperature $T_{c} = \pi
n_{s}/(2m_{H})$\cite{Kosterlitz}, where only coupled vortices are
present. Using the expressions (\ref{nn00}) and~(\ref{nn55}) for
the density $n_{s}$ of the superfluid component, we obtain an
equation for the Kosterlitz-Thouless transition temperature
$T_{c}$. Its solution is
%-------------------------------------------------------------------------
\begin{eqnarray}
\label{tct} T_c = \left[\left( 1 +
\sqrt{\frac{32}{27}\left(\frac{m_{H} T_{c}^{0}}{\pi n'}\right)^{3}
+ 1} \right)^{1/3}   - \left( \sqrt{\frac{32}{27}
\left(\frac{m_{H} T_{c}^{0}}{\pi n'}\right)^{3} + 1} - 1
\right)^{1/3}\right] \frac{T_{c}^{0}}{ 2^{1/3}}   .
\end{eqnarray}
%-------------------------------------------------------------------------
Here $T_{c}^{0}$ is an auxiliary quantity, equal to the
temperature at which the superfluid density vanishes in the
mean-field approximation $n_{s}(T_{c}^{0}) = 0$,
%-------------------------------------------------------------------------
\begin{equation}
\label{tct0} T_c^0 = \left( \frac{2 \pi n' c_s^4 m_{H}}{3 \zeta
(3)} \right)^{1/3}  .
\end{equation}
%-------------------------------------------------------------------------
In Eqs.~(\ref{tct}) and (\ref{tct0}), $n'$ is
%-------------------------------------------------------------------------
\begin{eqnarray}
\label{nexx} n' = n  - \frac{n Q}{2m_{H}c_{s}^{2}} .
\end{eqnarray}
%-------------------------------------------------------------------------
Even though the expression similar to Eq.~(\ref{tct}) has been
already visualized earlier in
Ref.~[\onlinecite{Berman_Snoke_Coalson}], the result of this paper
is nontrivial because it takes into account the influence of the
strong magnetic field on the excitonic superfluidity using mapping
of the system with magnetic field onto the system without field by
replacing excitonic mass by magnetic mass.

\section{Discussion}
\label{disc}

The superfluid density $n_s$ and the temperature of the
Kosterlitz-Thouless transition $T_c$ to the superfluid state for
the ``pure'' system  at fixed exciton density $n$ decrease as the
magnetic mass $m_{H}$ increases\cite{Berman_Tsvetus} (we can see
that by putting $Q = 0$, which  makes in Ref.~(\ref{tct}) $n' = n$
and $c_{s} = \sqrt{\mu/m_{H}}$). Since in strong magnetic fields
at $D \gg r_{H}$ the exciton magnetic mass is $m_H \approx
D^{3}\epsilon/(e^{2}r_{H}^{4})$ \cite{Ruvinskiy}, the superfluid
density $n_s$ and the temperature of the Kosterlitz-Thouless
transition $T_c$ decrease with increase of the magnetic field $H$,
the parameters of random field $\alpha_{e}$, $\alpha_{h}$ and
$g_{i}$  (Figs.~1 and~2).  Since in the ``dirty'' systems, $n_{s}$
and $T_{c}$ decrease with the increase of effective random field
$Q$ (analogous to the case without magnetic field
Ref.[\onlinecite{Berman_Snoke_Coalson}]), and in the strong
magnetic field  $Q$ is proportional to $m_{H}$
(Eq.~(\ref{Q_res_ap})), the increase of the magnetic field $H$
increases  the effective renormalized random field $Q$, and
suppress the superfluid density $n_s$ and the temperature of the
Kosterlitz-Thouless transition $T_c$.

 Note that in  the
presence of the disorder there is a quantum transition to the
superfluid state at zero temperature $T=0$ with respect to the
magnetic field $H$ and the parameters of the disorder
$\alpha_{e}$, $\alpha_{h}$ and $g_{i}$ (Figs.~1 and~2). While in
the ``pure'' system at any magnetic field $H$ at  there is always
the region in the density-temperature space, where the
superfluidity occurs\cite{Berman_Tsvetus}, in the presence of the
disorder at sufficiently large magnetic field $H$ or the
parameters of the disorder $\alpha_{e}$, $\alpha_{h}$ and $g_{i}$
there is no superfluidity at any exciton density.

Note also that in magnetic field the superfluid density $n_s$ and
the temperature of the Kosterlitz-Thouless transition $T_c$
decrease when the separation between quantum wells $D$ increases
because $n_{s}$ and $T_{c}$ are decreasing functions of the
magnetoexciton effecive mass $m_{H}$ (Eqs.~(\ref{nn55}) and
~(\ref{tct})), which is an increasing function of
$D$\cite{Ruvinskiy}. The dependence of $n_{s}$ and $T_{c}$ on $D$
through magnetic mass $m_H \approx D^{3}\epsilon/(e^{2}r_{H}^{4})$
and through decreasing dependence on the random field  $Q$, which
increases with the rise of $m_{H}$ (Eq.~(\ref{Q_res_ap})), is
stronger than their increasing logarithmic dependence on $D$
through the velocity of sound related to the logarithmic
dependence on $D$ of the chemical potential of the dipole-dipole
repulsion (Eq.~(\ref{Mu}). The latter results in the increasing
dependence of $n_{s}$ and $T_{c}$ on $D$ in a case without
magnetic field.\cite{Berman_Snoke_Coalson} The decreasing
dependence of $n_{s}$ and $T_{c}$ on $D$ at high magnetic fields
$H$ is significant when $D$ is large ($D \sim 30 r_{H}$), which is
far from $D$ used in experiments, because at very large values of
$D$ the superfluid system of magnetoexitons must transform into
the system of two incompressible liquids. \cite{Yoshioka}

In the high magnetic field limit at high $D$, the effective random
field is not small, and the approaches assuming coupling with the
random field to be much smaller than the dipole-dipole
repulsion\cite{Berman_Ruvinsky,Huang,Huang_Meng,Lopatin} are not
applicable. Note that in the present work the parameter $Q/\mu$ is
not required to be sufficiently small, and our formulas for the
superfluid density and Kosterlitz-Thouless temperature can be used
in the regime of realistic experimental parameters
 taken from photoluminescence line broadening
 measurements\cite{Snoke_paper}. The
coherent potential approximation (CPA) allowed us to derive the
exciton Green's function for the wide range of the random field,
and in the weak-scattering limit CPA results in the second-order
Born approximation.

The system of ``dirty'' indirect magnetoexcitons  can appear also
in unbalanced two-layer {\it electron} system  in  CQW in strong
magnetic fields near the filling factor $\nu =
1$\cite{Berman_Tsvetus}. An external electric voltage between
quantum wells change the filling, so say in the first
 quantum well the filling factor will be
$\nu _{1} = \Delta \nu \ll \frac{1}{2}$ and in another one it will
be $\nu _{2} = 1 - \Delta \nu $. Unbalanced filling factors
 $\nu _{1} = 1 + \Delta \nu $,
 $\nu _{2} = 1 - \Delta \nu $ is also possible.
  Thus in the first
quantum well (QW) there are rare electrons on the second Landau
level, and in the second QW there are rare empty places (holes) in
the first Landau level.
  Excess electrons in the first QW and holes in the second
QW can bind to indirect magnetoexcitons with the density $n_{ex} =
eH\Delta \nu /2\pi $.   Superfluidity in two-layer $e-e$ system in
high magnetic fields in the cases mentioned is analogous to the
superfluidity of two-layer $e-h$ system\cite{Berman_Tsvetus}.

Note that in this paper we assumed spin degeneracy factor $s = 1$
in high magnetic fields $H$. Since at $H = 0$ the spin degeneracy
is  $s = 4$, \ \cite{Berman_Snoke_Coalson} at low $H$ lifting spin
degeneracy will have substantial effect on the magnetoexciton
superfluidity, which we do not take into account in this paper
where we consider only high $H$.

\section*{Acknowledgements}
We thank Drs. Valerii Vinokur, Alexei Lopatin (Argonne National
Laboratory), Profs.~Peter Littlewood and Roland Zimmermann and the
participants of 1st Int. Conf. on Spontaneous Coherence in
Excitonic Systems for many useful and stimulating discussions. Yu.
E.~L. was supported by the INTAS and RFBR grants. D. W.~S. and R.
D.~C. have been supported by the National Science Foundation.

\newpage

\newpage

\begin{figure}
\rotatebox{270}{
\includegraphics[width = 16cm, height = 17cm]{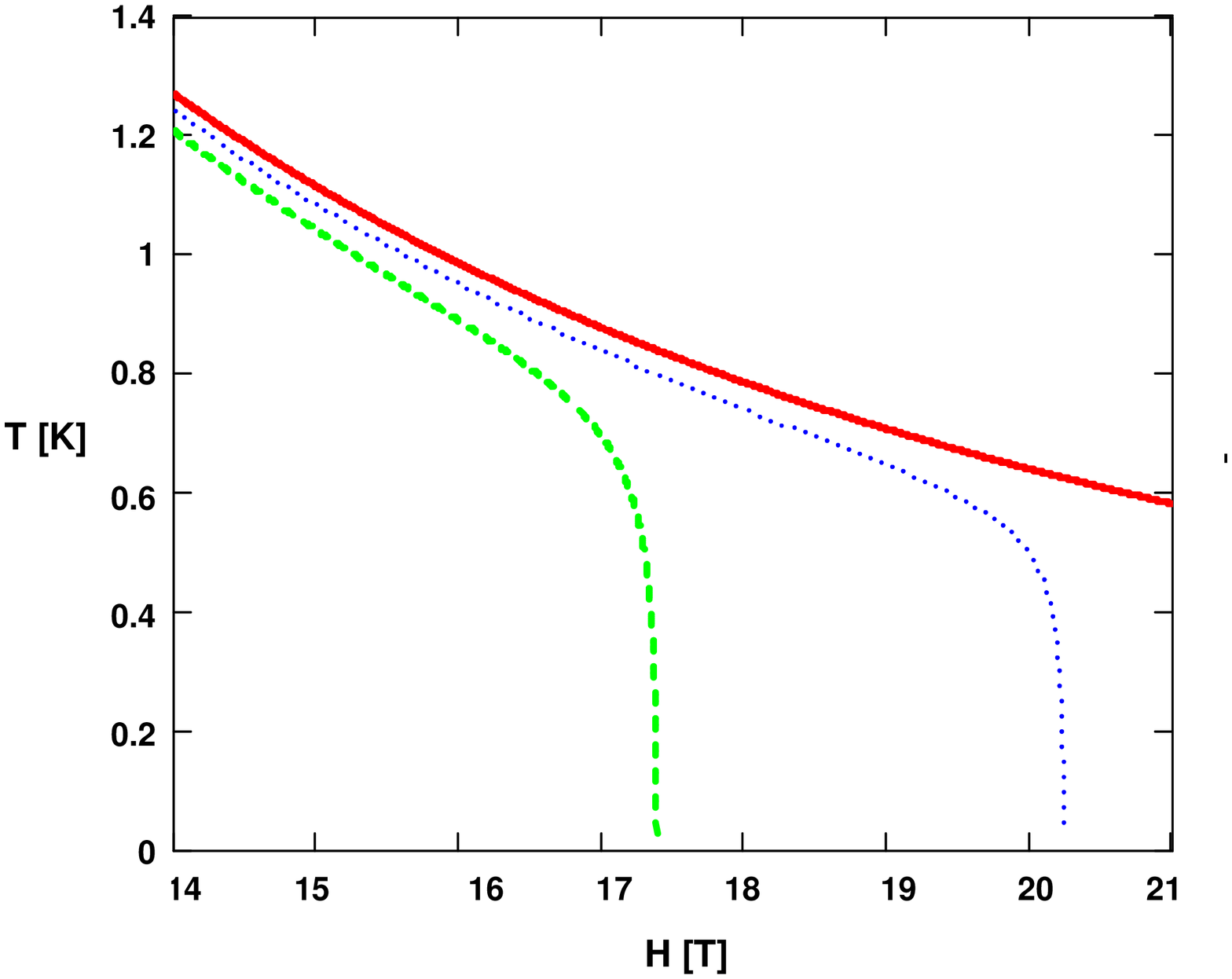}}
\caption{Dependence of temperature of Kosterlitz-Thouless
transition $T_c = T_c (H)$ (in units $K$; for $GaAs/GaAsAl$:
$\epsilon = 13$ on the magnetic field  $H$ (in units  $T$) at
$g_{i} = 10^{4}nm^{4}$ at the interwell distance $D = 15 nm$; at
the exciton density $n = 1.0\times 10^{11} cm^{-2}$; at the
different parameters of the random field $\alpha =\alpha_{e} =
\alpha_{h}$: $\alpha = 0$ -- solid curve; $\alpha = 0.5 meV/nm$ --
dotted curve; $\alpha = 0.7 meV/nm$ -- dashed curve.}
\end{figure}

\newpage

\begin{figure}
\rotatebox{270}{
\includegraphics[width = 16cm, height = 17cm]{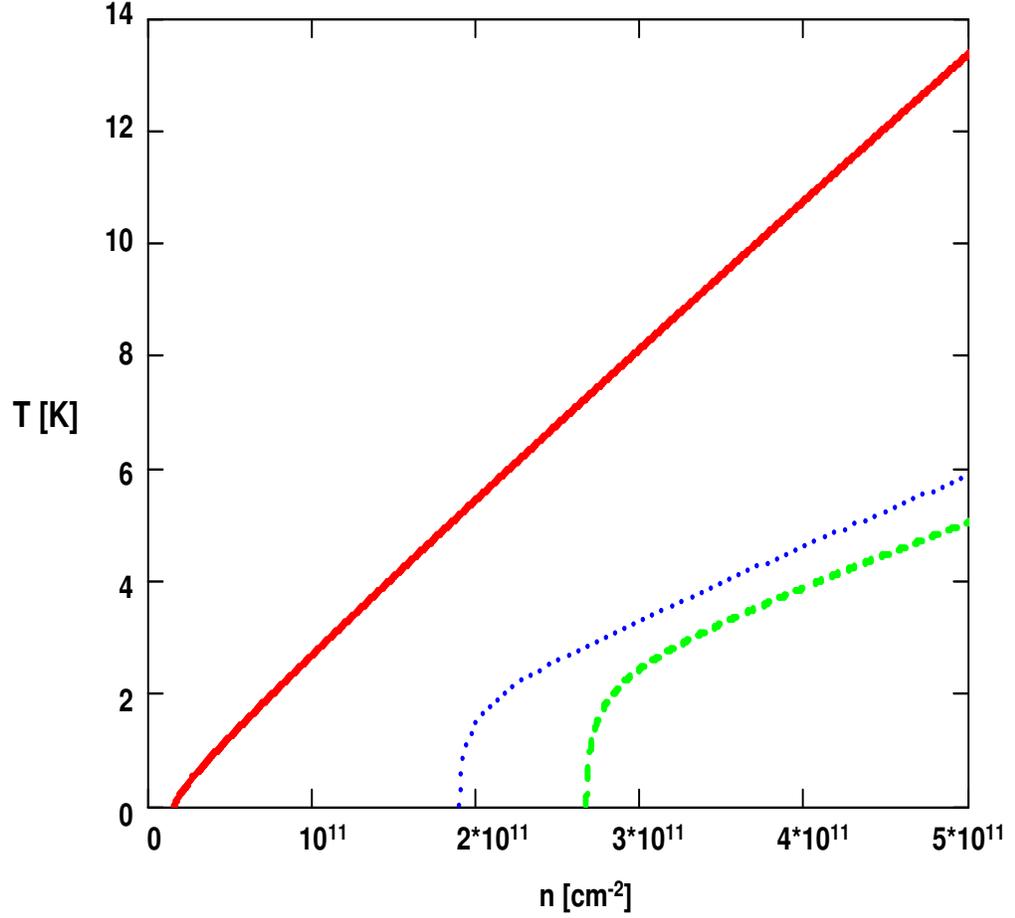}}
\caption{Dependence of temperature of Kosterlitz-Thouless
transition $T_{c} = T_{c}(n)$ (in units $K$; for $GaAs/GaAsAl$:
 $M = 0.24 m_{0}$; $\epsilon = 13$; $m_{0}$ is a mass of electron) on
the exciton density $n$ (in units $cm^{-2}$) at the interwell
distance $D = 15 nm$ at the parameters of the random field
$\alpha_{e} = \alpha_{h}= 1.5 meV/nm$; $g_{i} = 10^{4}nm^{4}$; at
different magnetic fields $H$ (in units of $T$): $H = 0$ -- solid
curve (for $H=0$ we use the results of Ref.[30] for spin
degeneracy factor $s=4$); $H = 14 T$ -- dotted curve; $H = 15 T$
-- dashed curve.}
\end{figure}

\end{document}